\documentclass[%
preprint,
nofootinbib,
 amsmath,amssymb,
 aps,
]{revtex4-2}

\usepackage{graphicx}
\usepackage{dcolumn}
\usepackage{bm}
\usepackage{comment}

\usepackage{xcolor}
\usepackage[normalem]{ulem}

\newcommand*\diff{\mathop{}\!\mathrm{d}}
\newcommand{\RomanNumeralCaps}[1]

\begin{document}

\title{A fundamental limit on energy savings in controlled channel flow, and how to beat it
}

\author{Daniel Floryan}\email{dfloryan@uh.edu}

\affiliation{
  Department of Mechanical Engineering, University of Houston, Houston, TX 77204, USA
}%

\date{\today}

\begin{abstract}
We derive a limit on energy savings in controlled channel flow. For flow in a channel driven by pressure, shear, or any combination of the two, and controlled via wall transpiration or spanwise wall motion, the uncontrolled laminar state requires the least net energy (accounting for the energetic cost of control). Thus, the optimal control solution is to laminarize the flow. Additionally, we raise the possibility of beating this limit. By simultaneously applying wall transpiration and spanwise wall motion, we show that it may be possible to attain sustained sub-laminar energy expenditure in a controlled flow. We provide a necessary design criterion for net energy savings. 
\end{abstract}

\maketitle

\section{Introduction}	
\label{sec:intro}	

In many flows of practical interest, a common goal is to reduce drag. This is because frictional drag is the main culprit constraining speed and efficiency, and also contributes to wear. Notable victims of drag include airplanes, ships, and fluid-carrying pipelines. Accordingly, significant effort has been---and is being---put forth to develop flow control strategies to reduce drag. 

Among the many forms of flow control, here we focus on two: transpiration (blowing/suction) and spanwise wall motion. A number of prior studies have demonstrated the ability to achieve drag reduction when using transpiration \citep{choi1994active, lee1997application, bewley2001dns, min2006sustained, quadrio2007effect, lieu2010controlling, mamori2014effect, gomez2016streamwise, koganezawa2019pathline, han2020active, park2020machine, jiao2021on, jiao2021use} or spanwise wall oscillations \citep{jung1992suppression, choi1998drag, choi2002drag, quadrio2004critical, ricco2008wall, quadrio2009streamwise, viotti2009streamwise, auteri2010experimental, yakeno2014modification, gatti2016reynolds, meysonnat2016experimental, bird2018experimental, skote2019wall, yao2019reynolds, marusic2021energy, ricco2021review}, even attaining sustained sub-laminar levels of drag \citep{min2006sustained, jiao2021on, jiao2021use}. Reducing drag saves energy that would otherwise be lost to friction, but the control input requires energy. Thus, despite reducing drag, controlling a flow may increase the energy expenditure on balance. 

It is worthwhile asking whether controlling a flow can confer a net energetic benefit, and whether there are any fundamental limits to how much one stands to gain. \citet{bewley2009fundamental} and \citet{fukagata2009lower} provide a partial answer. In their work, they prove that for pressure-driven flow, the energetic cost of transpiration is always greater than or equal to the energy saved due to drag reduction below the laminar level, for any distribution of transpiration (\citeauthor{bewley2009fundamental} showed this for channel flow, while \citeauthor{fukagata2009lower} showed this for a duct with arbitrary constant-shape cross-section and also included the effects of an arbitrary body force). In other words, no matter the spatiotemporal pattern of transpiration used, or level of drag reduction attained---even if sub-laminar---the uncontrolled laminar flow requires the least net energy. This is an important result: it rigorously establishes that the optimal control solution, from an energetic standpoint, is to laminarize the flow. (As an exception, \citet{fukagata2009lower} raise the possibility that transpiration in a duct with varying cross-sectional shape may reduce net energy requirements, although it has not yet been demonstrated.)

In this work, we generalize the result of \citet{bewley2009fundamental} and \citet{fukagata2009lower}, showing that the same fundamental limit on energy holds not only for pressure-driven flows, but also for shear-driven and mixed pressure- and shear-driven flows. We also show that the same fundamental limit exists when the control takes the form of arbitrary spanwise wall motion instead of transpiration. Finally, and perhaps most interestingly, we raise the possibility of beating this fundamental limit by combining transpiration with spanwise wall motion. That is, we show that it may be possible to attain sustained sub-laminar energy expenditure in a controlled flow.

\section{Derivation of a fundamental limit on energy savings}
\label{sec:deriv}

\begin{figure}
  \begin{center}
  \includegraphics[width=0.4\linewidth]{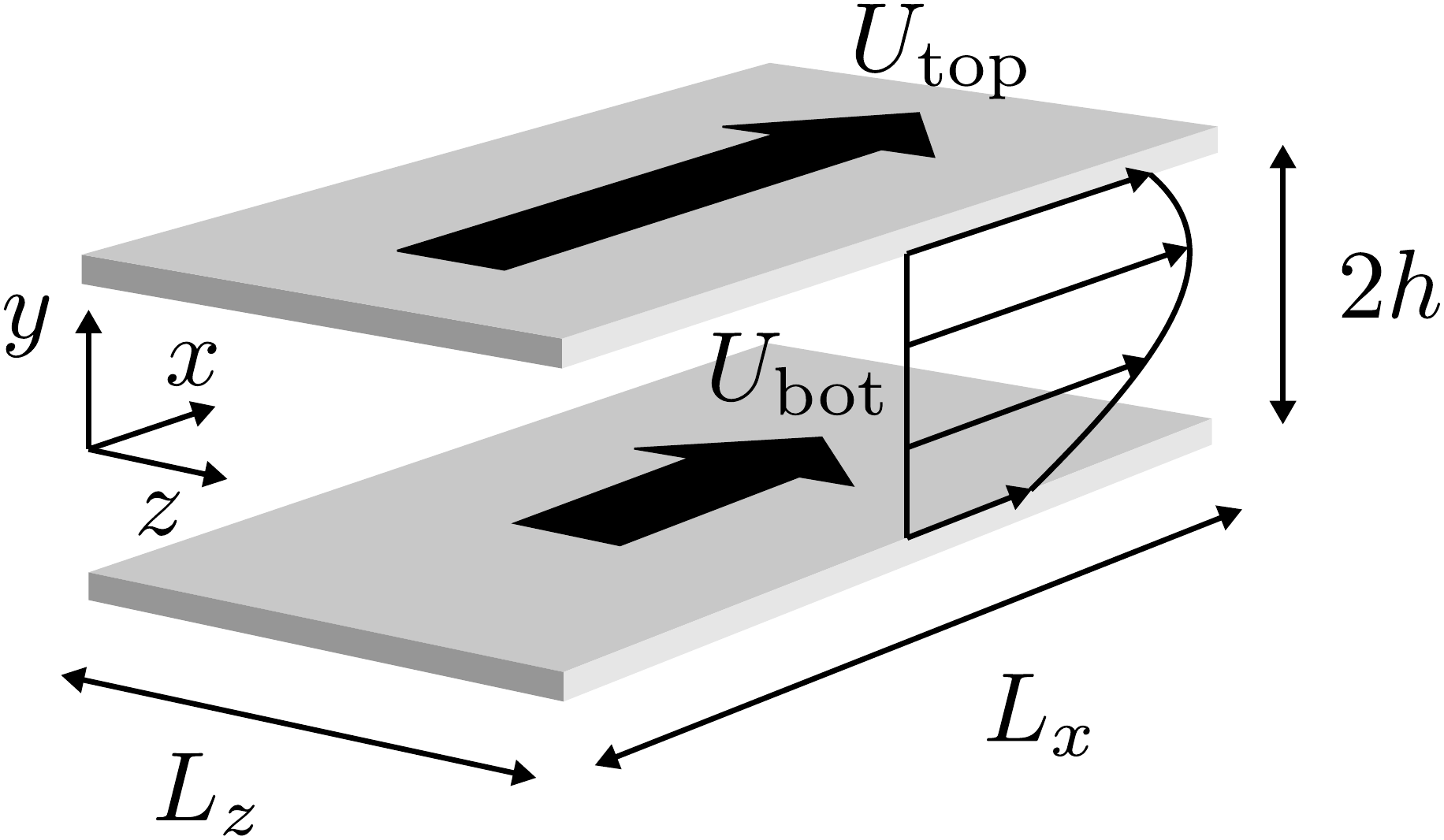}
  \end{center}
  \caption{Sketch of domain and uncontrolled laminar flow.}
  \label{fig:sketch}
\end{figure}

Consider constant-density flow in a straight channel bounded at the top and bottom by walls. The bottom wall moves with a constant velocity $U_{\text{bot}} \boldsymbol{i}$, the top wall moves with a constant velocity $U_{\text{top}} \boldsymbol{i}$, and we impose a pressure gradient $P_x \boldsymbol{i}$, where $\boldsymbol{i}$ is the unit vector in the streamwise direction. The flow satisfies the continuity and Navier-Stokes equations, 
\begin{align}
  \nabla \cdot \boldsymbol{u} &= 0, \label{eq:mass} \\
  \rho \left( \frac{\partial \boldsymbol{u}}{\partial t} + \boldsymbol{u} \cdot \nabla \boldsymbol{u} \right) &= -\nabla p + \mu \nabla^2 \boldsymbol{u} - P_x \boldsymbol{i}, \label{eq:mom}
\end{align}
on the domain $\Omega = \{ (x, y, z) \in [0, L_x] \times [-h, h] \times [0, L_z] \}$ with boundary $\partial \Omega$, sketched in Figure~\ref{fig:sketch}. Above, ${\boldsymbol{u} = (u, v, w)}$ is the velocity field, $p$ is the pressure field, $\rho$ is the density, and $\mu$ is the dynamic viscosity. The pressure gradient is constant in space but may depend on time, adjusted such that the bulk velocity
\begin{equation}
  \label{eq:bulk}
  U_B = \frac{1}{2 L_x h L_z} \int_0^{L_x} \int_{-h}^{h} \int_0^{L_z} u \diff x \diff y \diff z
\end{equation}
is constant. The flow is periodic in the streamwise ($x$) and spanwise ($z$) directions. When ${U_{\text{bot}} = U_{\text{top}} = 0}$, we have pressure-driven Poiseuille flow. When $P_x = 0$, we have shear-driven Couette flow. Otherwise, we have a mixed Couette-Poiseuille flow driven by shear and pressure. 

At the walls, located at $y = \pm h$, we apply control in two forms: transpiration and spanwise wall motion. We allow these controls to have arbitrary spatial and temporal distributions. An implication of the flow being periodic in $x$ and $z$ is that the net mass flux through the walls must be zero. 

The uncontrolled laminar flow has a velocity field 
\begin{equation}
  \label{eq:ulam}
  \boldsymbol{u}_L = u_L \boldsymbol{i} = \left[ \frac{P_x (y^2 - h^2)}{2 \mu} + \frac{(U_{\text{top}} - U_{\text{bot}}) y}{2h} + \frac{U_{\text{top}} + U_{\text{bot}}}{2} \right] \boldsymbol{i}
\end{equation}
and a total pressure field $P_x x$. Work is done to maintain the bulk velocity and to move the walls against forces. The question we address is whether net energy can be saved relative to the uncontrolled laminar flow by controlling the flow at the walls, accounting for the energy expenditure of the control. 

Starting from~\eqref{eq:mom}, take an inner product with the velocity vector and integrate over the domain to arrive at
\begin{equation}
  \label{eq:main1}
  \frac{\rho}{2} \int_\Omega \frac{\partial}{\partial t} ( \boldsymbol{u} \cdot \boldsymbol{u} ) \diff V + \rho \int_\Omega \boldsymbol{u} \cdot (\boldsymbol{u} \cdot \nabla \boldsymbol{u} ) \diff V = - \int_\Omega \boldsymbol{u} \cdot \nabla p \diff V + \mu \int_\Omega \boldsymbol{u} \cdot \nabla^2 \boldsymbol{u} \diff V - \int_\Omega u P_x \diff V.
\end{equation}
The convective, pressure, and viscous terms can all be simplified, and we proceed to do so one by one. 

By applying vector identities, continuity, and the divergence theorem, the convective term can be rewritten as 
\begin{equation}
  \label{eq:conv}
  \int_\Omega \boldsymbol{u} \cdot (\boldsymbol{u} \cdot \nabla \boldsymbol{u} ) \diff V = \oint_{\partial \Omega} \frac{1}{2} (\boldsymbol{u} \cdot \boldsymbol{u}) (\boldsymbol{u} \cdot \boldsymbol{n}) \diff A.
\end{equation}

By applying a vector identity, continuity, and the divergence theorem, the pressure term can be rewritten as 
\begin{equation}
  \label{eq:pres}
  \int_\Omega \boldsymbol{u} \cdot \nabla p \diff V = \oint_{\partial \Omega} p \boldsymbol{u} \cdot \boldsymbol{n} \diff A.
\end{equation}

By applying vector identities and the divergence theorem, the viscous term can be rewritten as 
\begin{equation}
  \label{eq:visc}
  \int_\Omega \boldsymbol{u} \cdot \nabla^2 \boldsymbol{u} \diff V = \oint_{\partial \Omega} \boldsymbol{u} \cdot (\boldsymbol{n} \cdot \nabla \boldsymbol{u}) \diff A - \int_\Omega \nabla \boldsymbol{u} \boldsymbol{:} \nabla \boldsymbol{u} \diff V.
\end{equation}

With these simplifications,~\eqref{eq:main1} becomes
\begin{equation}
\begin{aligned}
  \label{eq:main2}
  & \frac{\rho}{2} \int_\Omega \frac{\partial}{\partial t} ( \boldsymbol{u} \cdot \boldsymbol{u} ) \diff V + \frac{\rho}{2} \oint_{\partial \Omega} (\boldsymbol{u} \cdot \boldsymbol{u}) (\boldsymbol{u} \cdot \boldsymbol{n}) \diff A \\
  =& - \oint_{\partial \Omega} p \boldsymbol{u} \cdot \boldsymbol{n} \diff A + \mu \oint_{\partial \Omega} \boldsymbol{u} \cdot (\boldsymbol{n} \cdot \nabla \boldsymbol{u}) \diff A - \mu \int_\Omega \nabla \boldsymbol{u} \boldsymbol{:} \nabla \boldsymbol{u} \diff V - \int_\Omega u P_x \diff V,
\end{aligned}
\end{equation}
which can be written as
\begin{equation}
\begin{aligned}
  \label{eq:main3}
  &\frac{\diff}{\diff t} \frac{\rho}{2} \Vert \boldsymbol{u} \Vert^2 + \mu \Vert \nabla \boldsymbol{u} \Vert^2 + \oint_{\partial \Omega} \left( p + \frac{\rho}{2} \boldsymbol{u} \cdot \boldsymbol{u} \right) \boldsymbol{u} \cdot \boldsymbol{n} \diff A + 2 L_x h L_z U_B P_x \\
  =& \mu \oint_{\partial \Omega} \boldsymbol{u} \cdot (\boldsymbol{n} \cdot \nabla \boldsymbol{u}) \diff A,
\end{aligned}
\end{equation}
with the norms for vector and tensor fields defined implicitly. Due to the periodic boundary conditions, only the quantities at the walls contribute to the integrals along the boundary $\partial \Omega$. Therefore, \eqref{eq:main3} simplifies to
\begin{equation}
\begin{aligned}
  \label{eq:main4}
  &\frac{\diff}{\diff t} \frac{\rho}{2} \Vert \boldsymbol{u} \Vert^2 + \mu \Vert \nabla \boldsymbol{u} \Vert^2 + \int_{\text{walls}} \left( p + \frac{\rho}{2} \boldsymbol{u} \cdot \boldsymbol{u} \right) \boldsymbol{u} \cdot \boldsymbol{n} \diff A + 2 L_x h L_z U_B P_x \\
  =& \mu \int_{\text{walls}} \boldsymbol{u} \cdot (\boldsymbol{n} \cdot \nabla \boldsymbol{u}) \diff A.
\end{aligned}
\end{equation}

In order to maintain the motion of the walls, forces must be applied to them to balance friction and momentum flux due to transpiration. Performing a control volume analysis on the top and bottom walls reveals that the required force on each wall is
\begin{equation}
  \label{eq:ftop}
  \boldsymbol{F}_{\text{top/bot}} = \int_{\text{top/bot}} \rho \boldsymbol{u} (\boldsymbol{u} \cdot \boldsymbol{n}_i) \diff A - \int_{\text{top/bot}} \mu (\nabla \boldsymbol{u} + \nabla \boldsymbol{u}^\text{T}) \cdot \boldsymbol{n}_i \diff A.
\end{equation}
Note that we have written the unit normal as $\boldsymbol{n}_i$ to distinguish it from the unit normal used previously; they are related by $\boldsymbol{n}_i = - \boldsymbol{n}$. The first term is due to momentum flux from transpiration, and the second term is due to viscous forces. 

Next, we define an effective traction vector
\begin{equation}
  \label{eq:eftrac}
  \boldsymbol{t} := \rho \boldsymbol{u} (\boldsymbol{u} \cdot \boldsymbol{n}_i) - \mu (\nabla \boldsymbol{u} + \nabla \boldsymbol{u}^\text{T}) \cdot \boldsymbol{n}_i.
\end{equation}
We rewrite the forces on the walls as
\begin{equation}
  \boldsymbol{F}_{\text{top/bot}} = \int_{\text{top/bot}} \boldsymbol{t} \diff A.
\end{equation}
The rate of work done in order to move the walls is
\begin{equation}
  \dot W_{\text{top/bot}} = \int_{\text{top/bot}} \boldsymbol{u}_{\text{wall}} \cdot \boldsymbol{t} \diff A.
\end{equation}
Note that $\boldsymbol{u}_{\text{wall}}$ differs from the fluid velocity at the walls by
\begin{equation}
  \label{eq:uwall}
  \boldsymbol{u}_{\text{wall}} = \boldsymbol{u} - (\boldsymbol{u} \cdot \boldsymbol{n}) \boldsymbol{n}.
\end{equation}
It does not include the velocity component normal to the walls so that $\dot W_{\text{top}}$ and $\dot W_{\text{bot}}$ only account for the rate of work associated with the motion of the walls themselves. After some algebraic manipulation, the rate of work done in order to maintain both walls' motions, $\dot W_{\text{walls}} = \dot W_{\text{top}} + \dot W_{\text{bot}}$, is
\begin{align}
  \dot W_{\text{walls}} &= \int_{\text{walls}} \left[ \rho ( \boldsymbol{u}_{\text{wall}} \cdot \boldsymbol{u} ) ( \boldsymbol{u} \cdot \boldsymbol{n}_i ) - \mu \boldsymbol{n}_i \cdot (\boldsymbol{u}_{\text{wall}} \cdot \nabla \boldsymbol{u} ) - \mu \boldsymbol{u}_{\text{wall}} \cdot ( \boldsymbol{n}_i \cdot \nabla \boldsymbol{u} ) \right] \diff A. \label{eq:wtop}
\end{align}

Subtracting~\eqref{eq:wtop} from~\eqref{eq:main4} and rearranging terms yields
\begin{equation}
\begin{aligned}
  \label{eq:main5}
  &\frac{\diff}{\diff t} \frac{\rho}{2} \Vert \boldsymbol{u} \Vert^2 + \mu \Vert \nabla \boldsymbol{u} \Vert^2 + \int_{\text{walls}} \left( p + \frac{\rho}{2} \boldsymbol{u} \cdot \boldsymbol{u} \right) \boldsymbol{u} \cdot \boldsymbol{n} \diff A + 2 L_x h L_z U_B P_x \\
  =& \rho \int_{\text{walls}} (\boldsymbol{u}_{\text{wall}} \cdot \boldsymbol{u}) (\boldsymbol{u} \cdot \boldsymbol{n}) \diff A - \mu \int_{\text{walls}} \boldsymbol{n} \cdot (\boldsymbol{u}_{\text{wall}} \cdot \nabla \boldsymbol{u}) \diff A + \dot W_{\text{walls}} \\
  &+ \mu \int_{\text{walls}} (\boldsymbol{u} \cdot \boldsymbol{n}) \boldsymbol{n} \cdot (\boldsymbol{n} \cdot \nabla \boldsymbol{u}) \diff A.
\end{aligned}
\end{equation}
Note that $\boldsymbol{u}_{\text{wall}} \cdot \boldsymbol{u} = \boldsymbol{u}_{\text{wall}} \cdot \boldsymbol{u}_{\text{wall}}$. 

Now take the time-average of~\eqref{eq:main5}, where the time-average of a function $f$ is defined by
\begin{equation}
  \label{eq:tave}
  \langle f \rangle := \lim_{T \rightarrow \infty} \frac{1}{T} \int_0^T f(t) \diff t.
\end{equation}
Assuming $\Vert \boldsymbol{u} \Vert^2$ remains bounded, time-averaging yields
\begin{equation}
\begin{aligned}
  \label{eq:main6}
  & \mu \langle \Vert \nabla \boldsymbol{u} \Vert^2 \rangle + \left \langle \int_{\text{walls}} \left( p + \frac{\rho}{2} \boldsymbol{u} \cdot \boldsymbol{u} \right) \boldsymbol{u} \cdot \boldsymbol{n} \diff A \right \rangle + 2 L_x h L_z U_B \langle P_x \rangle \\
  =& \rho \left \langle \int_{\text{walls}} (\boldsymbol{u}_{\text{wall}} \cdot \boldsymbol{u}_{\text{wall}}) (\boldsymbol{u} \cdot \boldsymbol{n}) \diff A \right \rangle - \mu \left \langle \int_{\text{walls}} \boldsymbol{n} \cdot (\boldsymbol{u}_{\text{wall}} \cdot \nabla \boldsymbol{u}) \diff A \right \rangle + \langle \dot W_{\text{walls}} \rangle \\
  & + \mu \left \langle \int_{\text{walls}} (\boldsymbol{u} \cdot \boldsymbol{n}) \boldsymbol{n} \cdot (\boldsymbol{n} \cdot \nabla \boldsymbol{u}) \diff A \right \rangle.
\end{aligned}
\end{equation}

The norm of the velocity gradient can be rewritten. Let $\boldsymbol{u} = \boldsymbol{u}_L + \boldsymbol{u}'$, where $\boldsymbol{u}_L$ is the uncontrolled laminar flow from~\eqref{eq:ulam}, and $\boldsymbol{u}' = (u', v', w')$ is the deviation from it. Substituting this decomposition into the expression for the norm of the velocity gradient gives
\begin{equation}
  \label{eq:14}
  \Vert \nabla \boldsymbol{u} \Vert^2 = \Vert \nabla \boldsymbol{u}_L \Vert^2 + \Vert \nabla \boldsymbol{u}' \Vert^2 + 2 \int_\Omega \nabla \boldsymbol{u}_L \boldsymbol{:} \nabla \boldsymbol{u}' \diff V.
\end{equation}
Since the uncontrolled laminar flow has only one component, and it is a function of only $y$, the last term can be rewritten as
\begin{equation}
  \label{eq:cross}
  \int_\Omega \nabla \boldsymbol{u}_L \boldsymbol{:} \nabla \boldsymbol{u}' \diff V = - \frac{\diff^2 u_L}{\diff y^2} \int_\Omega u' \diff V +  \int_0^{L_z} \int_0^{L_x} \frac{\diff u_L}{\diff y} u' \Bigr|_{y = -h}^{h} \diff x \diff z,
\end{equation}
where we have used integration by parts and the fact that $u_L$ is quadratic in $y$. The first term is zero since the uncontrolled laminar flow and the controlled flow have the same bulk flow in the $x$ direction. The second term is zero since $u' \equiv 0$ on the walls (because the uncontrolled laminar flow and controlled flow have the same boundary condition for the $x$ component), making the entire expression in~\eqref{eq:cross} equal to zero. Thus, the norm of the velocity gradient is equal to the sum of that of the uncontrolled laminar flow and that of the deviation from the uncontrolled laminar flow; in this sense, the two velocity gradients may be thought of as being orthogonal. Substituting~\eqref{eq:14}--\eqref{eq:cross} into~\eqref{eq:main6} gives
\begin{equation}
\begin{aligned}
  \label{eq:main7}
  & \mu \Vert \nabla \boldsymbol{u}_L \Vert^2 + \mu \langle \Vert \nabla \boldsymbol{u}' \Vert^2 \rangle +  \left \langle \int_{\text{walls}} \left( p + \frac{\rho}{2} \boldsymbol{u} \cdot \boldsymbol{u} \right) \boldsymbol{u} \cdot \boldsymbol{n} \diff A \right \rangle  + 2 L_x h L_z U_B \langle P_x \rangle \\
  =& \rho \left \langle \int_{\text{walls}} (\boldsymbol{u}_{\text{wall}} \cdot \boldsymbol{u}_{\text{wall}}) (\boldsymbol{u} \cdot \boldsymbol{n}) \diff A \right \rangle - \mu \left \langle \int_{\text{walls}} \boldsymbol{n} \cdot (\boldsymbol{u}_{\text{wall}} \cdot \nabla \boldsymbol{u}) \diff A \right \rangle + \langle \dot W_{\text{walls}} \rangle \\
  & + \mu \left \langle \int_{\text{walls}} (\boldsymbol{u} \cdot \boldsymbol{n}) \boldsymbol{n} \cdot (\boldsymbol{n} \cdot \nabla \boldsymbol{u}) \diff A \right \rangle.
\end{aligned}
\end{equation}

Applying~\eqref{eq:main7} to the uncontrolled laminar flow yields
\begin{equation}
  \label{eq:lam}
  \mu \Vert \nabla \boldsymbol{u}_L \Vert^2 + 2 L_x h L_z U_B P_{x,L} = \dot W_{\text{walls},L},
\end{equation}
where a subscript $L$ denotes laminar quantities. The third, fifth, and last terms in~\eqref{eq:main7} do not appear in~\eqref{eq:lam} because there is no flow normal to the walls when there is no control; the sixth term does not appear because the laminar flow is rectilinear. Subtracting~\eqref{eq:lam} from~\eqref{eq:main7} gives
\begin{equation}
\begin{aligned}
  \label{eq:main8}
  & \mu \langle \Vert \nabla \boldsymbol{u}' \Vert^2 \rangle +  \left \langle \int_{\text{walls}} \left( p + \frac{\rho}{2} \boldsymbol{u} \cdot \boldsymbol{u} \right) \boldsymbol{u} \cdot \boldsymbol{n} \diff A \right \rangle  + 2 L_x h L_z U_B ( \langle P_x \rangle - P_{x,L} ) \\
  =& \rho \left \langle \int_{\text{walls}} (\boldsymbol{u}_{\text{wall}} \cdot \boldsymbol{u}_{\text{wall}}) (\boldsymbol{u} \cdot \boldsymbol{n}) \diff A \right \rangle - \mu \left \langle \int_{\text{walls}} \boldsymbol{n} \cdot (\boldsymbol{u}_{\text{wall}} \cdot \nabla \boldsymbol{u}) \diff A \right \rangle \\
  &+ \langle \dot W_{\text{walls}} \rangle - \dot W_{\text{walls},L} + \mu \left \langle \int_{\text{walls}} (\boldsymbol{u} \cdot \boldsymbol{n}) \boldsymbol{n} \cdot (\boldsymbol{n} \cdot \nabla \boldsymbol{u}) \diff A \right \rangle.
\end{aligned}
\end{equation}

Some of the terms in~\eqref{eq:main8} can be simplified. Putting ourselves in the frame of reference where $U_\text{bot} = -U_\text{top}$, it follows that $\boldsymbol{u}_{\text{wall}} \cdot \boldsymbol{u}_{\text{wall}} = U_\text{top}^2 + w^2$, where $w$ is the spanwise component of the velocity, and the integral in the fourth term becomes
\begin{equation}
  U_\text{top}^2 \int_{\text{walls}} (\boldsymbol{u} \cdot \boldsymbol{n}) \diff A + \int_{\text{walls}} w^2 (\boldsymbol{u} \cdot \boldsymbol{n}) \diff A = \int_{\text{walls}} w^2 (\boldsymbol{u} \cdot \boldsymbol{n}) \diff A,
\end{equation}
since the net mass flux through the walls must be zero. 
The fifth term in~\eqref{eq:main8} simplifies to
\begin{equation}
  \label{eq:last}
  \int_{\text{walls}} \boldsymbol{n} \cdot (\boldsymbol{u}_{\text{wall}} \cdot \nabla \boldsymbol{u}) \diff A = \int_0^{L_z} \int_0^{L_x} v \frac{\partial v}{\partial y} \Bigr|_{y = -h}^{h} \diff x \diff z,
\end{equation}
where we have used integration by parts, continuity, and periodicity of the flow in $x$ and $z$. The last term in~\eqref{eq:main8} simplifies to
\begin{equation}
  \label{eq:last2}
  \int_{\text{walls}} (\boldsymbol{u} \cdot \boldsymbol{n}) \boldsymbol{n} \cdot (\boldsymbol{n} \cdot \nabla \boldsymbol{u}) \diff A = \int_0^{L_z} \int_0^{L_x} v \frac{\partial v}{\partial y} \Bigr|_{y = -h}^h \diff x \diff z,
\end{equation}
which is the same as in~\eqref{eq:last}. These two terms cancel, and~\eqref{eq:main8} simplifies to
\begin{equation}
\begin{aligned}
  \label{eq:main9}
  & \mu \langle \Vert \nabla \boldsymbol{u}' \Vert^2 \rangle +  \left \langle \int_{\text{walls}} \left( p + \frac{\rho}{2} \boldsymbol{u} \cdot \boldsymbol{u} \right) \boldsymbol{u} \cdot \boldsymbol{n} \diff A \right \rangle  + 2 L_x h L_z U_B ( \langle P_x \rangle - P_{x,L} ) \\
  =& \rho \left \langle \int_{\text{walls}} w^2 (\boldsymbol{u} \cdot \boldsymbol{n}) \diff A \right \rangle + \langle \dot W_{\text{walls}} \rangle - \dot W_{\text{walls},L}.
\end{aligned}
\end{equation}

Finally, we rearrange terms to arrive at
\begin{equation}
\begin{aligned}
  \label{eq:main10}
  & - 2 L_x h L_z U_B \langle P_x \rangle + \langle \dot W_{\text{walls}} \rangle - \left \langle \int_{\text{walls}} \left( p + \frac{\rho}{2} \boldsymbol{u} \cdot \boldsymbol{u} \right) \boldsymbol{u} \cdot \boldsymbol{n} \diff A \right \rangle \\
  & - \left( - 2L_x h L_z U_B P_{x,L} + \dot W_{\text{walls},L} \right) \\
  =& \mu \langle \Vert \nabla \boldsymbol{u}' \Vert^2 \rangle - \rho \left \langle \int_{\text{walls}} w^2 (\boldsymbol{u} \cdot \boldsymbol{n}) \diff A \right \rangle.
\end{aligned}
\end{equation}
The left-hand side is the difference in power needed to maintain the controlled flow (first line) and the uncontrolled laminar flow (second line). This includes the power needed to apply the pressure gradient, the power needed to maintain the motion of the walls, and the power needed to apply transpiration, which includes the rate of work done against pressure and the rate of injection of kinetic energy into the flow. 

Note that~\eqref{eq:main10} also holds for open channel flow, as in the simulations by \citet{marusic2021energy}, so our ensuing discussion may also hold some relevance for boundary layer flows, as in the accompanying experiments of \citet{marusic2021energy}. However, we have not proven that such a relation holds for boundary layer flows. Indeed, the relevance to boundary layer flows is complicated by the fact that they are spatially developing (see \citet{ricco2022integral} for an example of how spatially developing flows may fundamentally differ in this type of study).

\section{Discussion}
\label{sec:disc}

Suppose we use only transpiration for control, as in the analyses of \citet{bewley2009fundamental} and \citet{fukagata2009lower}. Then $w \equiv 0$ at the walls, and the right-hand side of~\eqref{eq:main10} is equal to $\mu \langle \Vert \nabla \boldsymbol{u}' \Vert^2 \rangle \ge 0$. Thus, the left-hand side, which is the difference in power needed to maintain the controlled flow and the uncontrolled laminar flow, is non-negative. In other words, the uncontrolled laminar flow requires the least net energy no matter the spatiotemporal distribution of the control. From an energetic standpoint, the optimal solution is to laminarize the flow. This holds whether the flow is driven by pressure, shear, or any combination of the two, recovering the result of \citet{bewley2009fundamental} and \citet{fukagata2009lower} as a special case. 

Now suppose that we use only spanwise wall motion for control. Then $\boldsymbol{u} \cdot \boldsymbol{n} = 0$ at the walls, and the right-hand side of~\eqref{eq:main10} is equal to $\mu \langle \Vert \nabla \boldsymbol{u}' \Vert^2 \rangle \ge 0$. We conclude again that the uncontrolled laminar flow requires the least net energy no matter the spatiotemporal distribution of the control. 

Finally, consider the case where we simultaneously apply transpiration and spanwise motion at the walls. In this case, the last term in~\eqref{eq:main10} is non-zero. We may interpret this term as the covariance between the square of the spanwise speed and the transpiration speed, that is, as the covariance between the two forms of control. Physically, this term originates from the work done on the walls to maintain their motions. Specifically, it is the work associated with the component of the external force that arises due to the momentum flux across the walls (the first terms on the right-hand sides of~\eqref{eq:ftop} and~\eqref{eq:eftrac}). If this covariance is greater than the increase in the norm of the velocity gradient due to the controlled flow, then the controlled flow requires less net power than the uncontrolled laminar flow. Mathematically, the criterion for net energy savings relative to the uncontrolled laminar flow is 
\begin{equation}
  \label{eq:sav}
  \rho \left \langle \int_{\text{walls}} w^2 (\boldsymbol{u} \cdot \boldsymbol{n}) \diff A \right \rangle > \mu \langle \Vert \nabla \boldsymbol{u}' \Vert^2 \rangle.
\end{equation}
Physically, this criterion states that if the negative of the average work arising due to momentum flux is greater than the average additional spatial variation in the flow induced by control, then the sustained net energy expenditure is sub-laminar.

Sustained sub-laminar drag has previously been attained by passive means \citep{mohammadi2013pressure}, implying sub-laminar energy expenditure. 
We are unaware, however, of any results demonstrating net energy savings relative to the uncontrolled laminar flow when using active flow control, making this possibility rather interesting (\citet{fukagata2009lower} raised the possibility of doing so by applying transpiration around a bump on a channel wall, but it was never demonstrated). Moreover, the criterion in~\eqref{eq:sav} is constructive since the covariance term can be completely specified as it only contains control terms. For example, this criterion reveals a necessary condition for net energy savings: spanwise wall motion and transpiration must have spatiotemporal overlap. In particular, the spanwise wall motion must, on average, be greater in regions of suction than in regions of blowing. This condition is not sufficient, however, since it is unknown \textit{a priori} whether the designed control induces additional dissipation greater than the covariance term. 

Nevertheless, the criterion provides some insight. The increase in the norm of the velocity gradient induced by control contributes to increased dissipation relative to the uncontrolled flow. Since dissipation tends to be greatest near walls, the additional dissipation induced by the control will be dominated by contributions in the vicinity of the control. With both terms in~\eqref{eq:sav} depending on near-wall or wall quantities, the criterion provides a path forward for rational design of the control. 

It is important to note that the criterion in~\eqref{eq:sav} is ideal in the sense that it considers all of the negative work to be recoverable. In a real system, the amount of negative work that can be recovered depends on the devices used to implement control. This is an important consideration in any physical flow control system. Nevertheless, our theoretical findings open the possibility of sustained sub-laminar energy expenditure, an important first step. Pursuing this possibility is certainly worthy of future work.


\textit{Acknowledgements:} The author thanks Drs J. M. Floryan and A. Prosperetti for helpful feedback. 


\textit{Declaration of interests:} The author reports no conflict of interest.


\textit{Author ORCID:} Daniel Floryan, https://orcid.org/0000-0002-6353-5075.


\bibliography{references}

\begin{thebibliography}{33}%
\makeatletter
\providecommand \@ifxundefined [1]{%
 \@ifx{#1\undefined}
}%
\providecommand \@ifnum [1]{%
 \ifnum #1\expandafter \@firstoftwo
 \else \expandafter \@secondoftwo
 \fi
}%
\providecommand \@ifx [1]{%
 \ifx #1\expandafter \@firstoftwo
 \else \expandafter \@secondoftwo
 \fi
}%
\providecommand \natexlab [1]{#1}%
\providecommand \enquote  [1]{``#1''}%
\providecommand \bibnamefont  [1]{#1}%
\providecommand \bibfnamefont [1]{#1}%
\providecommand \citenamefont [1]{#1}%
\providecommand \href@noop [0]{\@secondoftwo}%
\providecommand \href [0]{\begingroup \@sanitize@url \@href}%
\providecommand \@href[1]{\@@startlink{#1}\@@href}%
\providecommand \@@href[1]{\endgroup#1\@@endlink}%
\providecommand \@sanitize@url [0]{\catcode `\\12\catcode `\$12\catcode
  `\&12\catcode `\#12\catcode `\^12\catcode `\_12\catcode `\%12\relax}%
\providecommand \@@startlink[1]{}%
\providecommand \@@endlink[0]{}%
\providecommand \url  [0]{\begingroup\@sanitize@url \@url }%
\providecommand \@url [1]{\endgroup\@href {#1}{\urlprefix }}%
\providecommand \urlprefix  [0]{URL }%
\providecommand \Eprint [0]{\href }%
\providecommand \doibase [0]{https://doi.org/}%
\providecommand \selectlanguage [0]{\@gobble}%
\providecommand \bibinfo  [0]{\@secondoftwo}%
\providecommand \bibfield  [0]{\@secondoftwo}%
\providecommand \translation [1]{[#1]}%
\providecommand \BibitemOpen [0]{}%
\providecommand \bibitemStop [0]{}%
\providecommand \bibitemNoStop [0]{.\EOS\space}%
\providecommand \EOS [0]{\spacefactor3000\relax}%
\providecommand \BibitemShut  [1]{\csname bibitem#1\endcsname}%
\let\auto@bib@innerbib\@empty
\bibitem [{\citenamefont {Choi}\ \emph {et~al.}(1994)\citenamefont {Choi},
  \citenamefont {Moin},\ and\ \citenamefont {Kim}}]{choi1994active}%
  \BibitemOpen
  \bibfield  {author} {\bibinfo {author} {\bibfnamefont {H.}~\bibnamefont
  {Choi}}, \bibinfo {author} {\bibfnamefont {P.}~\bibnamefont {Moin}},\ and\
  \bibinfo {author} {\bibfnamefont {J.}~\bibnamefont {Kim}},\ }\bibfield
  {title} {\bibinfo {title} {Active turbulence control for drag reduction in
  wall-bounded flows},\ }\href@noop {} {\bibfield  {journal} {\bibinfo
  {journal} {Journal of Fluid Mechanics}\ }\textbf {\bibinfo {volume} {262}},\
  \bibinfo {pages} {75} (\bibinfo {year} {1994})}\BibitemShut {NoStop}%
\bibitem [{\citenamefont {Lee}\ \emph {et~al.}(1997)\citenamefont {Lee},
  \citenamefont {Kim}, \citenamefont {Babcock},\ and\ \citenamefont
  {Goodman}}]{lee1997application}%
  \BibitemOpen
  \bibfield  {author} {\bibinfo {author} {\bibfnamefont {C.}~\bibnamefont
  {Lee}}, \bibinfo {author} {\bibfnamefont {J.}~\bibnamefont {Kim}}, \bibinfo
  {author} {\bibfnamefont {D.}~\bibnamefont {Babcock}},\ and\ \bibinfo {author}
  {\bibfnamefont {R.}~\bibnamefont {Goodman}},\ }\bibfield  {title} {\bibinfo
  {title} {Application of neural networks to turbulence control for drag
  reduction},\ }\href@noop {} {\bibfield  {journal} {\bibinfo  {journal}
  {Physics of Fluids}\ }\textbf {\bibinfo {volume} {9}},\ \bibinfo {pages}
  {1740} (\bibinfo {year} {1997})}\BibitemShut {NoStop}%
\bibitem [{\citenamefont {Bewley}\ \emph {et~al.}(2001)\citenamefont {Bewley},
  \citenamefont {Moin},\ and\ \citenamefont {Temam}}]{bewley2001dns}%
  \BibitemOpen
  \bibfield  {author} {\bibinfo {author} {\bibfnamefont {T.~R.}\ \bibnamefont
  {Bewley}}, \bibinfo {author} {\bibfnamefont {P.}~\bibnamefont {Moin}},\ and\
  \bibinfo {author} {\bibfnamefont {R.}~\bibnamefont {Temam}},\ }\bibfield
  {title} {\bibinfo {title} {{DNS}-based predictive control of turbulence: an
  optimal benchmark for feedback algorithms},\ }\href@noop {} {\bibfield
  {journal} {\bibinfo  {journal} {Journal of Fluid Mechanics}\ }\textbf
  {\bibinfo {volume} {447}},\ \bibinfo {pages} {179} (\bibinfo {year}
  {2001})}\BibitemShut {NoStop}%
\bibitem [{\citenamefont {Min}\ \emph {et~al.}(2006)\citenamefont {Min},
  \citenamefont {Kang}, \citenamefont {Speyer},\ and\ \citenamefont
  {Kim}}]{min2006sustained}%
  \BibitemOpen
  \bibfield  {author} {\bibinfo {author} {\bibfnamefont {T.}~\bibnamefont
  {Min}}, \bibinfo {author} {\bibfnamefont {S.~M.}\ \bibnamefont {Kang}},
  \bibinfo {author} {\bibfnamefont {J.~L.}\ \bibnamefont {Speyer}},\ and\
  \bibinfo {author} {\bibfnamefont {J.}~\bibnamefont {Kim}},\ }\bibfield
  {title} {\bibinfo {title} {Sustained sub-laminar drag in a fully developed
  channel flow},\ }\href@noop {} {\bibfield  {journal} {\bibinfo  {journal}
  {Journal of Fluid Mechanics}\ }\textbf {\bibinfo {volume} {558}},\ \bibinfo
  {pages} {309} (\bibinfo {year} {2006})}\BibitemShut {NoStop}%
\bibitem [{\citenamefont {Quadrio}\ \emph {et~al.}(2007)\citenamefont
  {Quadrio}, \citenamefont {Floryan},\ and\ \citenamefont
  {Luchini}}]{quadrio2007effect}%
  \BibitemOpen
  \bibfield  {author} {\bibinfo {author} {\bibfnamefont {M.}~\bibnamefont
  {Quadrio}}, \bibinfo {author} {\bibfnamefont {J.~M.}\ \bibnamefont
  {Floryan}},\ and\ \bibinfo {author} {\bibfnamefont {P.}~\bibnamefont
  {Luchini}},\ }\bibfield  {title} {\bibinfo {title} {Effect of
  streamwise-periodic wall transpiration on turbulent friction drag},\
  }\href@noop {} {\bibfield  {journal} {\bibinfo  {journal} {Journal of Fluid
  Mechanics}\ }\textbf {\bibinfo {volume} {576}},\ \bibinfo {pages} {425}
  (\bibinfo {year} {2007})}\BibitemShut {NoStop}%
\bibitem [{\citenamefont {Lieu}\ \emph {et~al.}(2010)\citenamefont {Lieu},
  \citenamefont {Moarref},\ and\ \citenamefont
  {Jovanovi{\'c}}}]{lieu2010controlling}%
  \BibitemOpen
  \bibfield  {author} {\bibinfo {author} {\bibfnamefont {B.~K.}\ \bibnamefont
  {Lieu}}, \bibinfo {author} {\bibfnamefont {R.}~\bibnamefont {Moarref}},\ and\
  \bibinfo {author} {\bibfnamefont {M.~R.}\ \bibnamefont {Jovanovi{\'c}}},\
  }\bibfield  {title} {\bibinfo {title} {Controlling the onset of turbulence by
  streamwise travelling waves. part 2. direct numerical simulation},\
  }\href@noop {} {\bibfield  {journal} {\bibinfo  {journal} {Journal of Fluid
  Mechanics}\ }\textbf {\bibinfo {volume} {663}},\ \bibinfo {pages} {100}
  (\bibinfo {year} {2010})}\BibitemShut {NoStop}%
\bibitem [{\citenamefont {Mamori}\ \emph {et~al.}(2014)\citenamefont {Mamori},
  \citenamefont {Iwamoto},\ and\ \citenamefont {Murata}}]{mamori2014effect}%
  \BibitemOpen
  \bibfield  {author} {\bibinfo {author} {\bibfnamefont {H.}~\bibnamefont
  {Mamori}}, \bibinfo {author} {\bibfnamefont {K.}~\bibnamefont {Iwamoto}},\
  and\ \bibinfo {author} {\bibfnamefont {A.}~\bibnamefont {Murata}},\
  }\bibfield  {title} {\bibinfo {title} {Effect of the parameters of traveling
  waves created by blowing and suction on the relaminarization phenomena in
  fully developed turbulent channel flow},\ }\href@noop {} {\bibfield
  {journal} {\bibinfo  {journal} {Physics of Fluids}\ }\textbf {\bibinfo
  {volume} {26}},\ \bibinfo {pages} {015101} (\bibinfo {year}
  {2014})}\BibitemShut {NoStop}%
\bibitem [{\citenamefont {G{\'o}mez}\ \emph {et~al.}(2016)\citenamefont
  {G{\'o}mez}, \citenamefont {Blackburn}, \citenamefont {Rudman}, \citenamefont
  {Sharma},\ and\ \citenamefont {McKeon}}]{gomez2016streamwise}%
  \BibitemOpen
  \bibfield  {author} {\bibinfo {author} {\bibfnamefont {F.}~\bibnamefont
  {G{\'o}mez}}, \bibinfo {author} {\bibfnamefont {H.~M.}\ \bibnamefont
  {Blackburn}}, \bibinfo {author} {\bibfnamefont {M.}~\bibnamefont {Rudman}},
  \bibinfo {author} {\bibfnamefont {A.~S.}\ \bibnamefont {Sharma}},\ and\
  \bibinfo {author} {\bibfnamefont {B.~J.}\ \bibnamefont {McKeon}},\ }\bibfield
   {title} {\bibinfo {title} {Streamwise-varying steady transpiration control
  in turbulent pipe flow},\ }\href@noop {} {\bibfield  {journal} {\bibinfo
  {journal} {Journal of Fluid Mechanics}\ }\textbf {\bibinfo {volume} {796}},\
  \bibinfo {pages} {588} (\bibinfo {year} {2016})}\BibitemShut {NoStop}%
\bibitem [{\citenamefont {Koganezawa}\ \emph {et~al.}(2019)\citenamefont
  {Koganezawa}, \citenamefont {Mitsuishi}, \citenamefont {Shimura},
  \citenamefont {Iwamoto}, \citenamefont {Mamori},\ and\ \citenamefont
  {Murata}}]{koganezawa2019pathline}%
  \BibitemOpen
  \bibfield  {author} {\bibinfo {author} {\bibfnamefont {S.}~\bibnamefont
  {Koganezawa}}, \bibinfo {author} {\bibfnamefont {A.}~\bibnamefont
  {Mitsuishi}}, \bibinfo {author} {\bibfnamefont {T.}~\bibnamefont {Shimura}},
  \bibinfo {author} {\bibfnamefont {K.}~\bibnamefont {Iwamoto}}, \bibinfo
  {author} {\bibfnamefont {H.}~\bibnamefont {Mamori}},\ and\ \bibinfo {author}
  {\bibfnamefont {A.}~\bibnamefont {Murata}},\ }\bibfield  {title} {\bibinfo
  {title} {Pathline analysis of traveling wavy blowing and suction control in
  turbulent pipe flow for drag reduction},\ }\href@noop {} {\bibfield
  {journal} {\bibinfo  {journal} {International Journal of Heat and Fluid
  Flow}\ }\textbf {\bibinfo {volume} {77}},\ \bibinfo {pages} {388} (\bibinfo
  {year} {2019})}\BibitemShut {NoStop}%
\bibitem [{\citenamefont {Han}\ and\ \citenamefont
  {Huang}(2020)}]{han2020active}%
  \BibitemOpen
  \bibfield  {author} {\bibinfo {author} {\bibfnamefont {B.-Z.}\ \bibnamefont
  {Han}}\ and\ \bibinfo {author} {\bibfnamefont {W.-X.}\ \bibnamefont
  {Huang}},\ }\bibfield  {title} {\bibinfo {title} {Active control for drag
  reduction of turbulent channel flow based on convolutional neural networks},\
  }\href@noop {} {\bibfield  {journal} {\bibinfo  {journal} {Physics of
  Fluids}\ }\textbf {\bibinfo {volume} {32}},\ \bibinfo {pages} {095108}
  (\bibinfo {year} {2020})}\BibitemShut {NoStop}%
\bibitem [{\citenamefont {Park}\ and\ \citenamefont
  {Choi}(2020)}]{park2020machine}%
  \BibitemOpen
  \bibfield  {author} {\bibinfo {author} {\bibfnamefont {J.}~\bibnamefont
  {Park}}\ and\ \bibinfo {author} {\bibfnamefont {H.}~\bibnamefont {Choi}},\
  }\bibfield  {title} {\bibinfo {title} {Machine-learning-based feedback
  control for drag reduction in a turbulent channel flow},\ }\href@noop {}
  {\bibfield  {journal} {\bibinfo  {journal} {Journal of Fluid Mechanics}\
  }\textbf {\bibinfo {volume} {904}} (\bibinfo {year} {2020})}\BibitemShut
  {NoStop}%
\bibitem [{\citenamefont {Jiao}\ and\ \citenamefont
  {Floryan}(2021{\natexlab{a}})}]{jiao2021on}%
  \BibitemOpen
  \bibfield  {author} {\bibinfo {author} {\bibfnamefont {L.}~\bibnamefont
  {Jiao}}\ and\ \bibinfo {author} {\bibfnamefont {J.~M.}\ \bibnamefont
  {Floryan}},\ }\bibfield  {title} {\bibinfo {title} {On the use of
  transpiration patterns for reduction of pressure losses},\ }\href@noop {}
  {\bibfield  {journal} {\bibinfo  {journal} {Journal of Fluid Mechanics}\
  }\textbf {\bibinfo {volume} {915}} (\bibinfo {year}
  {2021}{\natexlab{a}})}\BibitemShut {NoStop}%
\bibitem [{\citenamefont {Jiao}\ and\ \citenamefont
  {Floryan}(2021{\natexlab{b}})}]{jiao2021use}%
  \BibitemOpen
  \bibfield  {author} {\bibinfo {author} {\bibfnamefont {L.}~\bibnamefont
  {Jiao}}\ and\ \bibinfo {author} {\bibfnamefont {J.~M.}\ \bibnamefont
  {Floryan}},\ }\bibfield  {title} {\bibinfo {title} {Use of transpiration for
  reduction of resistance to relative movement of parallel plates},\
  }\href@noop {} {\bibfield  {journal} {\bibinfo  {journal} {Physical Review
  Fluids}\ }\textbf {\bibinfo {volume} {6}},\ \bibinfo {pages} {014101}
  (\bibinfo {year} {2021}{\natexlab{b}})}\BibitemShut {NoStop}%
\bibitem [{\citenamefont {Jung}\ \emph {et~al.}(1992)\citenamefont {Jung},
  \citenamefont {Mangiavacchi},\ and\ \citenamefont
  {Akhavan}}]{jung1992suppression}%
  \BibitemOpen
  \bibfield  {author} {\bibinfo {author} {\bibfnamefont {W.~J.}\ \bibnamefont
  {Jung}}, \bibinfo {author} {\bibfnamefont {N.}~\bibnamefont {Mangiavacchi}},\
  and\ \bibinfo {author} {\bibfnamefont {R.}~\bibnamefont {Akhavan}},\
  }\bibfield  {title} {\bibinfo {title} {Suppression of turbulence in
  wall-bounded flows by high-frequency spanwise oscillations},\ }\href@noop {}
  {\bibfield  {journal} {\bibinfo  {journal} {Physics of Fluids A: Fluid
  Dynamics}\ }\textbf {\bibinfo {volume} {4}},\ \bibinfo {pages} {1605}
  (\bibinfo {year} {1992})}\BibitemShut {NoStop}%
\bibitem [{\citenamefont {Choi}\ and\ \citenamefont
  {Graham}(1998)}]{choi1998drag}%
  \BibitemOpen
  \bibfield  {author} {\bibinfo {author} {\bibfnamefont {K.-S.}\ \bibnamefont
  {Choi}}\ and\ \bibinfo {author} {\bibfnamefont {M.}~\bibnamefont {Graham}},\
  }\bibfield  {title} {\bibinfo {title} {Drag reduction of turbulent pipe flows
  by circular-wall oscillation},\ }\href@noop {} {\bibfield  {journal}
  {\bibinfo  {journal} {Physics of Fluids}\ }\textbf {\bibinfo {volume} {10}},\
  \bibinfo {pages} {7} (\bibinfo {year} {1998})}\BibitemShut {NoStop}%
\bibitem [{\citenamefont {Choi}\ \emph {et~al.}(2002)\citenamefont {Choi},
  \citenamefont {Xu},\ and\ \citenamefont {Sung}}]{choi2002drag}%
  \BibitemOpen
  \bibfield  {author} {\bibinfo {author} {\bibfnamefont {J.-I.}\ \bibnamefont
  {Choi}}, \bibinfo {author} {\bibfnamefont {C.-X.}\ \bibnamefont {Xu}},\ and\
  \bibinfo {author} {\bibfnamefont {H.~J.}\ \bibnamefont {Sung}},\ }\bibfield
  {title} {\bibinfo {title} {Drag reduction by spanwise wall oscillation in
  wall-bounded turbulent flows},\ }\href@noop {} {\bibfield  {journal}
  {\bibinfo  {journal} {AIAA Journal}\ }\textbf {\bibinfo {volume} {40}},\
  \bibinfo {pages} {842} (\bibinfo {year} {2002})}\BibitemShut {NoStop}%
\bibitem [{\citenamefont {Quadrio}\ and\ \citenamefont
  {Ricco}(2004)}]{quadrio2004critical}%
  \BibitemOpen
  \bibfield  {author} {\bibinfo {author} {\bibfnamefont {M.}~\bibnamefont
  {Quadrio}}\ and\ \bibinfo {author} {\bibfnamefont {P.}~\bibnamefont
  {Ricco}},\ }\bibfield  {title} {\bibinfo {title} {Critical assessment of
  turbulent drag reduction through spanwise wall oscillations},\ }\href@noop {}
  {\bibfield  {journal} {\bibinfo  {journal} {Journal of Fluid Mechanics}\
  }\textbf {\bibinfo {volume} {521}},\ \bibinfo {pages} {251} (\bibinfo {year}
  {2004})}\BibitemShut {NoStop}%
\bibitem [{\citenamefont {Ricco}\ and\ \citenamefont
  {Quadrio}(2008)}]{ricco2008wall}%
  \BibitemOpen
  \bibfield  {author} {\bibinfo {author} {\bibfnamefont {P.}~\bibnamefont
  {Ricco}}\ and\ \bibinfo {author} {\bibfnamefont {M.}~\bibnamefont
  {Quadrio}},\ }\bibfield  {title} {\bibinfo {title} {Wall-oscillation
  conditions for drag reduction in turbulent channel flow},\ }\href@noop {}
  {\bibfield  {journal} {\bibinfo  {journal} {International Journal of Heat and
  Fluid Flow}\ }\textbf {\bibinfo {volume} {29}},\ \bibinfo {pages} {891}
  (\bibinfo {year} {2008})}\BibitemShut {NoStop}%
\bibitem [{\citenamefont {Quadrio}\ \emph {et~al.}(2009)\citenamefont
  {Quadrio}, \citenamefont {Ricco},\ and\ \citenamefont
  {Viotti}}]{quadrio2009streamwise}%
  \BibitemOpen
  \bibfield  {author} {\bibinfo {author} {\bibfnamefont {M.}~\bibnamefont
  {Quadrio}}, \bibinfo {author} {\bibfnamefont {P.}~\bibnamefont {Ricco}},\
  and\ \bibinfo {author} {\bibfnamefont {C.}~\bibnamefont {Viotti}},\
  }\bibfield  {title} {\bibinfo {title} {Streamwise-travelling waves of
  spanwise wall velocity for turbulent drag reduction},\ }\href@noop {}
  {\bibfield  {journal} {\bibinfo  {journal} {Journal of Fluid Mechanics}\
  }\textbf {\bibinfo {volume} {627}},\ \bibinfo {pages} {161} (\bibinfo {year}
  {2009})}\BibitemShut {NoStop}%
\bibitem [{\citenamefont {Viotti}\ \emph {et~al.}(2009)\citenamefont {Viotti},
  \citenamefont {Quadrio},\ and\ \citenamefont
  {Luchini}}]{viotti2009streamwise}%
  \BibitemOpen
  \bibfield  {author} {\bibinfo {author} {\bibfnamefont {C.}~\bibnamefont
  {Viotti}}, \bibinfo {author} {\bibfnamefont {M.}~\bibnamefont {Quadrio}},\
  and\ \bibinfo {author} {\bibfnamefont {P.}~\bibnamefont {Luchini}},\
  }\bibfield  {title} {\bibinfo {title} {Streamwise oscillation of spanwise
  velocity at the wall of a channel for turbulent drag reduction},\ }\href@noop
  {} {\bibfield  {journal} {\bibinfo  {journal} {Physics of Fluids}\ }\textbf
  {\bibinfo {volume} {21}},\ \bibinfo {pages} {115109} (\bibinfo {year}
  {2009})}\BibitemShut {NoStop}%
\bibitem [{\citenamefont {Auteri}\ \emph {et~al.}(2010)\citenamefont {Auteri},
  \citenamefont {Baron}, \citenamefont {Belan}, \citenamefont {Campanardi},\
  and\ \citenamefont {Quadrio}}]{auteri2010experimental}%
  \BibitemOpen
  \bibfield  {author} {\bibinfo {author} {\bibfnamefont {F.}~\bibnamefont
  {Auteri}}, \bibinfo {author} {\bibfnamefont {A.}~\bibnamefont {Baron}},
  \bibinfo {author} {\bibfnamefont {M.}~\bibnamefont {Belan}}, \bibinfo
  {author} {\bibfnamefont {G.}~\bibnamefont {Campanardi}},\ and\ \bibinfo
  {author} {\bibfnamefont {M.}~\bibnamefont {Quadrio}},\ }\bibfield  {title}
  {\bibinfo {title} {Experimental assessment of drag reduction by traveling
  waves in a turbulent pipe flow},\ }\href@noop {} {\bibfield  {journal}
  {\bibinfo  {journal} {Physics of Fluids}\ }\textbf {\bibinfo {volume} {22}},\
  \bibinfo {pages} {115103} (\bibinfo {year} {2010})}\BibitemShut {NoStop}%
\bibitem [{\citenamefont {Yakeno}\ \emph {et~al.}(2014)\citenamefont {Yakeno},
  \citenamefont {Hasegawa},\ and\ \citenamefont
  {Kasagi}}]{yakeno2014modification}%
  \BibitemOpen
  \bibfield  {author} {\bibinfo {author} {\bibfnamefont {A.}~\bibnamefont
  {Yakeno}}, \bibinfo {author} {\bibfnamefont {Y.}~\bibnamefont {Hasegawa}},\
  and\ \bibinfo {author} {\bibfnamefont {N.}~\bibnamefont {Kasagi}},\
  }\bibfield  {title} {\bibinfo {title} {Modification of quasi-streamwise
  vortical structure in a drag-reduced turbulent channel flow with spanwise
  wall oscillation},\ }\href@noop {} {\bibfield  {journal} {\bibinfo  {journal}
  {Physics of Fluids}\ }\textbf {\bibinfo {volume} {26}},\ \bibinfo {pages}
  {085109} (\bibinfo {year} {2014})}\BibitemShut {NoStop}%
\bibitem [{\citenamefont {Gatti}\ and\ \citenamefont
  {Quadrio}(2016)}]{gatti2016reynolds}%
  \BibitemOpen
  \bibfield  {author} {\bibinfo {author} {\bibfnamefont {D.}~\bibnamefont
  {Gatti}}\ and\ \bibinfo {author} {\bibfnamefont {M.}~\bibnamefont
  {Quadrio}},\ }\bibfield  {title} {\bibinfo {title} {Reynolds-number
  dependence of turbulent skin-friction drag reduction induced by spanwise
  forcing},\ }\href@noop {} {\bibfield  {journal} {\bibinfo  {journal} {Journal
  of Fluid Mechanics}\ }\textbf {\bibinfo {volume} {802}},\ \bibinfo {pages}
  {553} (\bibinfo {year} {2016})}\BibitemShut {NoStop}%
\bibitem [{\citenamefont {Meysonnat}\ \emph {et~al.}(2016)\citenamefont
  {Meysonnat}, \citenamefont {Roggenkamp}, \citenamefont {Li}, \citenamefont
  {Roidl},\ and\ \citenamefont {Schr{\"o}der}}]{meysonnat2016experimental}%
  \BibitemOpen
  \bibfield  {author} {\bibinfo {author} {\bibfnamefont {P.~S.}\ \bibnamefont
  {Meysonnat}}, \bibinfo {author} {\bibfnamefont {D.}~\bibnamefont
  {Roggenkamp}}, \bibinfo {author} {\bibfnamefont {W.}~\bibnamefont {Li}},
  \bibinfo {author} {\bibfnamefont {B.}~\bibnamefont {Roidl}},\ and\ \bibinfo
  {author} {\bibfnamefont {W.}~\bibnamefont {Schr{\"o}der}},\ }\bibfield
  {title} {\bibinfo {title} {Experimental and numerical investigation of
  transversal traveling surface waves for drag reduction},\ }\href@noop {}
  {\bibfield  {journal} {\bibinfo  {journal} {European Journal of Mechanics
  B/Fluids}\ }\textbf {\bibinfo {volume} {55}},\ \bibinfo {pages} {313}
  (\bibinfo {year} {2016})}\BibitemShut {NoStop}%
\bibitem [{\citenamefont {Bird}\ \emph {et~al.}(2018)\citenamefont {Bird},
  \citenamefont {Santer},\ and\ \citenamefont
  {Morrison}}]{bird2018experimental}%
  \BibitemOpen
  \bibfield  {author} {\bibinfo {author} {\bibfnamefont {J.}~\bibnamefont
  {Bird}}, \bibinfo {author} {\bibfnamefont {M.}~\bibnamefont {Santer}},\ and\
  \bibinfo {author} {\bibfnamefont {J.~F.}\ \bibnamefont {Morrison}},\
  }\bibfield  {title} {\bibinfo {title} {Experimental control of turbulent
  boundary layers with in-plane travelling waves},\ }\href@noop {} {\bibfield
  {journal} {\bibinfo  {journal} {Flow, Turbulence and Combustion}\ }\textbf
  {\bibinfo {volume} {100}},\ \bibinfo {pages} {1015} (\bibinfo {year}
  {2018})}\BibitemShut {NoStop}%
\bibitem [{\citenamefont {Skote}\ \emph {et~al.}(2019)\citenamefont {Skote},
  \citenamefont {Mishra},\ and\ \citenamefont {Wu}}]{skote2019wall}%
  \BibitemOpen
  \bibfield  {author} {\bibinfo {author} {\bibfnamefont {M.}~\bibnamefont
  {Skote}}, \bibinfo {author} {\bibfnamefont {M.}~\bibnamefont {Mishra}},\ and\
  \bibinfo {author} {\bibfnamefont {Y.}~\bibnamefont {Wu}},\ }\bibfield
  {title} {\bibinfo {title} {Wall oscillation induced drag reduction zone in a
  turbulent boundary layer},\ }\href@noop {} {\bibfield  {journal} {\bibinfo
  {journal} {Flow, Turbulence and Combustion}\ }\textbf {\bibinfo {volume}
  {102}},\ \bibinfo {pages} {641} (\bibinfo {year} {2019})}\BibitemShut
  {NoStop}%
\bibitem [{\citenamefont {Yao}\ \emph {et~al.}(2019)\citenamefont {Yao},
  \citenamefont {Chen},\ and\ \citenamefont {Hussain}}]{yao2019reynolds}%
  \BibitemOpen
  \bibfield  {author} {\bibinfo {author} {\bibfnamefont {J.}~\bibnamefont
  {Yao}}, \bibinfo {author} {\bibfnamefont {X.}~\bibnamefont {Chen}},\ and\
  \bibinfo {author} {\bibfnamefont {F.}~\bibnamefont {Hussain}},\ }\bibfield
  {title} {\bibinfo {title} {Reynolds number effect on drag control via
  spanwise wall oscillation in turbulent channel flows},\ }\href@noop {}
  {\bibfield  {journal} {\bibinfo  {journal} {Physics of Fluids}\ }\textbf
  {\bibinfo {volume} {31}},\ \bibinfo {pages} {085108} (\bibinfo {year}
  {2019})}\BibitemShut {NoStop}%
\bibitem [{\citenamefont {Marusic}\ \emph {et~al.}(2021)\citenamefont
  {Marusic}, \citenamefont {Chandran}, \citenamefont {Rouhi}, \citenamefont
  {Fu}, \citenamefont {Wine}, \citenamefont {Holloway}, \citenamefont {Chung},\
  and\ \citenamefont {Smits}}]{marusic2021energy}%
  \BibitemOpen
  \bibfield  {author} {\bibinfo {author} {\bibfnamefont {I.}~\bibnamefont
  {Marusic}}, \bibinfo {author} {\bibfnamefont {D.}~\bibnamefont {Chandran}},
  \bibinfo {author} {\bibfnamefont {A.}~\bibnamefont {Rouhi}}, \bibinfo
  {author} {\bibfnamefont {M.~K.}\ \bibnamefont {Fu}}, \bibinfo {author}
  {\bibfnamefont {D.}~\bibnamefont {Wine}}, \bibinfo {author} {\bibfnamefont
  {B.}~\bibnamefont {Holloway}}, \bibinfo {author} {\bibfnamefont
  {D.}~\bibnamefont {Chung}},\ and\ \bibinfo {author} {\bibfnamefont {A.~J.}\
  \bibnamefont {Smits}},\ }\bibfield  {title} {\bibinfo {title} {An
  energy-efficient pathway to turbulent drag reduction},\ }\href@noop {}
  {\bibfield  {journal} {\bibinfo  {journal} {Nature Communications}\ }\textbf
  {\bibinfo {volume} {12}},\ \bibinfo {pages} {1} (\bibinfo {year}
  {2021})}\BibitemShut {NoStop}%
\bibitem [{\citenamefont {Ricco}\ \emph {et~al.}(2021)\citenamefont {Ricco},
  \citenamefont {Skote},\ and\ \citenamefont {Leschziner}}]{ricco2021review}%
  \BibitemOpen
  \bibfield  {author} {\bibinfo {author} {\bibfnamefont {P.}~\bibnamefont
  {Ricco}}, \bibinfo {author} {\bibfnamefont {M.}~\bibnamefont {Skote}},\ and\
  \bibinfo {author} {\bibfnamefont {M.~A.}\ \bibnamefont {Leschziner}},\
  }\bibfield  {title} {\bibinfo {title} {A review of turbulent skin-friction
  drag reduction by near-wall transverse forcing},\ }\href@noop {} {\bibfield
  {journal} {\bibinfo  {journal} {Progress in Aerospace Sciences}\ }\textbf
  {\bibinfo {volume} {123}},\ \bibinfo {pages} {100713} (\bibinfo {year}
  {2021})}\BibitemShut {NoStop}%
\bibitem [{\citenamefont {Bewley}(2009)}]{bewley2009fundamental}%
  \BibitemOpen
  \bibfield  {author} {\bibinfo {author} {\bibfnamefont {T.~R.}\ \bibnamefont
  {Bewley}},\ }\bibfield  {title} {\bibinfo {title} {A fundamental limit on the
  balance of power in a transpiration-controlled channel flow},\ }\href@noop {}
  {\bibfield  {journal} {\bibinfo  {journal} {Journal of Fluid Mechanics}\
  }\textbf {\bibinfo {volume} {632}},\ \bibinfo {pages} {443} (\bibinfo {year}
  {2009})}\BibitemShut {NoStop}%
\bibitem [{\citenamefont {Fukagata}\ \emph {et~al.}(2009)\citenamefont
  {Fukagata}, \citenamefont {Sugiyama},\ and\ \citenamefont
  {Kasagi}}]{fukagata2009lower}%
  \BibitemOpen
  \bibfield  {author} {\bibinfo {author} {\bibfnamefont {K.}~\bibnamefont
  {Fukagata}}, \bibinfo {author} {\bibfnamefont {K.}~\bibnamefont {Sugiyama}},\
  and\ \bibinfo {author} {\bibfnamefont {N.}~\bibnamefont {Kasagi}},\
  }\bibfield  {title} {\bibinfo {title} {On the lower bound of net driving
  power in controlled duct flows},\ }\href@noop {} {\bibfield  {journal}
  {\bibinfo  {journal} {Physica D: Nonlinear Phenomena}\ }\textbf {\bibinfo
  {volume} {238}},\ \bibinfo {pages} {1082} (\bibinfo {year}
  {2009})}\BibitemShut {NoStop}%
\bibitem [{\citenamefont {Ricco}\ and\ \citenamefont
  {Skote}(2022)}]{ricco2022integral}%
  \BibitemOpen
  \bibfield  {author} {\bibinfo {author} {\bibfnamefont {P.}~\bibnamefont
  {Ricco}}\ and\ \bibinfo {author} {\bibfnamefont {M.}~\bibnamefont {Skote}},\
  }\bibfield  {title} {\bibinfo {title} {Integral relations for the
  skin-friction coefficient of canonical flows},\ }\href@noop {} {\bibfield
  {journal} {\bibinfo  {journal} {Journal of Fluid Mechanics}\ }\textbf
  {\bibinfo {volume} {943}},\ \bibinfo {pages} {A50} (\bibinfo {year}
  {2022})}\BibitemShut {NoStop}%
\bibitem [{\citenamefont {Mohammadi}\ and\ \citenamefont
  {Floryan}(2013)}]{mohammadi2013pressure}%
  \BibitemOpen
  \bibfield  {author} {\bibinfo {author} {\bibfnamefont {A.}~\bibnamefont
  {Mohammadi}}\ and\ \bibinfo {author} {\bibfnamefont {J.~M.}\ \bibnamefont
  {Floryan}},\ }\bibfield  {title} {\bibinfo {title} {Pressure losses in
  grooved channels},\ }\href@noop {} {\bibfield  {journal} {\bibinfo  {journal}
  {Journal of Fluid Mechanics}\ }\textbf {\bibinfo {volume} {725}},\ \bibinfo
  {pages} {23} (\bibinfo {year} {2013})}\BibitemShut {NoStop}%
\end{thebibliography}%

\end{document}